\pdfoutput=1
\documentclass[11pt,a4paper]{article}
\usepackage[T1]{fontenc}
\usepackage[utf8]{inputenc}
\usepackage{mathptmx}
\usepackage[scaled=0.92]{helvet}
\usepackage[margin=2.2cm]{geometry}
\usepackage{graphicx}
\usepackage{multirow}
\usepackage{amsmath,amssymb}
\usepackage{xcolor}
\usepackage{textcomp}
\usepackage{booktabs}
\usepackage{tabularx}
\usepackage{array}
\usepackage[font=small,labelfont=bf,labelsep=space]{caption}
\usepackage{microtype}
\usepackage{xurl}
\usepackage[hidelinks]{hyperref}

\newcommand{\repourl}{\url{https://github.com/securagenai/signed-but-steerable}}
\newcommand{\zenodourl}{archived release v1.0.2: \href{https://doi.org/10.5281/zenodo.23948839}{doi:10.5281/zenodo.23948839}}

\newcommand{\tablebodyfont}{\footnotesize}
\newcommand{\TCH}[1]{\textbf{#1}}
\newcommand{\botrule}{\bottomrule}
\newcommand{\bmhead}[1]{\par\medskip\noindent\textbf{#1.}\ }
\newcolumntype{L}[1]{>{\raggedright\arraybackslash}p{#1}}
\newcolumntype{C}[1]{>{\centering\arraybackslash}p{#1}}

\newcommand{\degr}{\ensuremath{^\circ}}

\title{\textbf{Signed but Steerable: Measuring and Closing the Text--Vector Integrity Gap in Offline Retrieval-Augmented Generation}}
\author{%
  {\fontencoding{OT1}\fontfamily{cmr}\selectfont Georgios Feretzakis and Alexandros Papaspyridis}\\[1em]
  {\normalsize\fontencoding{OT1}\fontfamily{cmr}\selectfont securagen.ai, Athens, Greece}\\
  {\normalsize\fontencoding{OT1}\selectfont\href{mailto:research@securagen.ai}{\textcolor{magenta}{\texttt{research@securagen.ai}}}}%
}
\date{}

\begin{document}
\maketitle

\begin{abstract}
Offline retrieval-augmented generation (RAG) on edge devices is updated through signed knowledge packs that carry text and precomputed embedding vectors. A signature proves who published a pack, not that its vectors encode its text. We study this text--vector integrity gap. A verifier that recomputes vectors and accepts them within a cosine tolerance concedes a predictable share of queries to an attacker who controls the vectors: the hijack rate equals the distribution of angular retrieval margins at the tolerance (the margin law). Margins are small in five corpora and two embedding models, so an idealised tolerance at the level of floating-point noise leaves up to 5.5\% of queries exposed (at most 1.2\% in any corpus once exact duplicate documents are excluded), and a threshold calibrated from our CPU INT8--FP32 comparison of a small embedding model (at the empirical 0.1\% quantile of honest cosines) leaves 98--100\% exposed; on a Raspberry Pi~5 running a local language model, such an attack lowers correct answers from 0.75 to 0.39. Integrity audits of compressed vector stores fare no better: on a FAISS product-quantisation index over 57,638 documents, tampered vectors that all pass the audit put the attacker's document first for 25\% of targets. Verifying quantisation codes instead, with ambiguity decided by the verifier, a calibrated guard band and canonicalisation, accepted every honest vector in our samples on three platforms and left the publisher no choice. We also show that common floating-point and dynamic-INT8 inference pipelines are not bit-reproducible across batch sizes or architectures, and report pre-registered negative results on abstention, context pruning and semantic screening. We release code and data.
\end{abstract}


\section{Introduction}\label{sec:intro}

Retrieval-augmented generation (RAG) lets a language model answer from a local corpus~\cite{r12}, and small quantised models now make fully offline RAG feasible on embedded boards~\cite{r9,r104}. In disconnected deployments, such as field, industrial and defence settings, knowledge reaches the device through update packages. Our earlier work proposed signed, versioned \emph{knowledge packs} for this purpose and evaluated them on a Raspberry Pi~5~\cite{prior}. It also left an open problem: a validly signed but poisoned pack passed every pack check.

This paper studies a sharper version of that problem. A knowledge pack typically ships text together with precomputed embedding vectors, because computing embeddings on a small device is slow; in our earlier study, embedding a 950-passage pack on the Pi's CPU took about 29~minutes~\cite{prior}. The signature then covers the vectors, but nothing guarantees that the vectors encode the text. An attacker who controls the publishing pipeline, or an insider, can ship benign text with vectors chosen to steer retrieval. The same gap exists in any vector database: what is stored is a vector, often compressed, and audits compare stored vectors with recomputed ones under some tolerance. Embedding weaknesses are recognised as a top risk for LLM applications~\cite{owasp25}, and per-embedding signatures have been proposed as a defence~\cite{vectorsmuggle}. We ask what such checks actually guarantee, and how to make them sound.

We address four research questions with experiments on a Raspberry Pi~5 running a local Qwen2.5-3B model, a desktop for scale, and an Apple M2 laptop as a third platform:
\begin{itemize}
\item \textbf{RQ1.} If a verifier accepts shipped vectors within a cosine tolerance of recomputed ones, how many queries can an attacker hijack, and can this be predicted without running attacks?
\item \textbf{RQ2.} How exposed are the compressed storage formats of vector databases to the same attack?
\item \textbf{RQ3.} Can exact verification close the gap across heterogeneous hardware, and what does it require?
\item \textbf{RQ4.} What does verification cost on the CPU of an edge device that also runs the language model, and what happens end to end?
\end{itemize}

Our contributions are:
\begin{enumerate}
\item \textbf{The margin law} (Sect.~\ref{sec:law}): for an optimal attacker, the hijack rate under tolerance $\tau$ equals the empirical distribution of angular retrieval margins at $\arccos\tau$. It matches attack simulation exactly and lets an operator measure exposure without attacking. On all answerable SQuAD~2.0 questions and four BEIR test sets (five corpora) with two embedding models, margins are small, so even float-noise tolerances leave measurable exposure, and a threshold calibrated from our CPU INT8--FP32 comparison leaves 98--100\% of queries exposed (Sect.~\ref{sec:res-law}).
\item \textbf{Exposure of compressed vector stores} (Sect.~\ref{sec:res-vdb}): tolerance-based audits against SQ8, SQ4, PQ and binary storage leave 2--15\%, 55--89\%, about 100\% and 100\% of queries exposed. On real FAISS indexes, an executed attack on product quantisation passes the tolerance check for every tampered vector and puts the attacker's document first for 25\% of targets; code-level verification blocks all of them.
\item \textbf{Exact code-level verification} (Sect.~\ref{sec:res-exact}): comparing quantisation codes, with ambiguity decided only by the verifier, a guard band calibrated to measured float noise and a verify-and-canonicalise rule, accepted every honest vector in our samples on three platforms and four compute backends, without special hardware. We also show that floating-point and dynamic-INT8 pipelines are not bit-reproducible across batch sizes or architectures.
\item \textbf{Measured on-device cost and end-to-end impact} (Sects.~\ref{sec:res-cost} and~\ref{sec:res-e2e}): verification on the device's CPU costs the concurrently running LLM 23--31\% of its decoding speed, the attack lowers the device's correct answers from 0.75 to 0.39 under the INT8 reference threshold, and layered pack verification rejects all twelve structural update attacks we tested.
\item \textbf{Pre-registered negative results} (Sect.~\ref{sec:res-neg}) on abstention, context pruning and semantic screening, and an open release of code, data and the protocol with its deviation log.
\end{enumerate}

\section{Background, Threat Model and Related Work}\label{sec:bg}

\subsection{Retrieval-Augmented Generation on Edge Devices}\label{sec:bg-rag}
A retrieval-augmented generation pipeline splits a corpus into chunks, maps every chunk to a dense vector with an embedding model, and stores the vectors in an index. At query time the query is embedded with the same model, the nearest chunks by cosine similarity (or inner product of normalised vectors) are retrieved, optionally reranked by a cross-encoder, and placed in the prompt of a language model~\cite{r12}. The quality and the safety of the answer therefore depend on which chunks the index returns, and that decision is taken entirely in vector space.

On edge devices the same pipeline runs without a network connection, with a small quantised language model~\cite{r9,r104} and memory- and latency-aware indexing~\cite{r16}. Knowledge reaches such a device through update packages. Embedding a corpus on a small board is slow, about 29~minutes for a 950-passage pack on the CPU of a Raspberry Pi~5 in our earlier study~\cite{prior}, so packs commonly ship precomputed vectors with the text. The device then has to decide whether to trust them.

\subsection{Vector Storage and Approximate Search}\label{sec:bg-vdb}
Vector databases rarely store full-precision vectors. To reduce memory and speed up search they store compressed codes: half-precision floats; scalar quantisation, which maps each dimension to one of $2^b$ levels between per-dimension bounds (SQ8, SQ4); product quantisation, which splits a vector into sub-vectors and stores, for each, the index of the nearest centroid in a learned codebook~\cite{pq}; anisotropic and other learned quantisers~\cite{guo2020}; or binary codes that keep only the sign of each dimension. These codes are combined with coarse partitioning or proximity graphs~\cite{r43} and are implemented in widely used libraries~\cite{faiss,johnson2021}. Every compressed format reconstructs a vector only approximately. The reconstruction error is small for half-precision floats, a fraction of a degree for SQ8, and tens of degrees for product and binary codes, as we measure in Sect.~\ref{sec:res-vdb}. Any audit that compares stored vectors with recomputed ones must tolerate this error.

\subsection{Knowledge Packs and Their Integrity}\label{sec:bg-packs}
A knowledge pack~\cite{prior} contains text chunks, a manifest with operations (add, supersede, delete), optional precomputed vectors, and signatures. Role separation in the spirit of TUF~\cite{r85} lets a content key add chunks while changes to existing chunks require a threshold of editor signatures; a transparency log in the spirit of Certificate Transparency~\cite{ct} and monotonic version state protect against rollback and against conflicting extensions of a device's retained history; manifests follow the IETF information model for IoT firmware updates~\cite{r86}. These mechanisms descend from software supply-chain security, where signatures, attestations and transparency logs establish who produced an artefact and how~\cite{sigstore,intoto2019}, and from recent specifications for signing machine-learning models~\cite{r83}. They bind bytes to keys and to a history. They do not bind a vector to the text it is supposed to encode: a valid signature over a pack whose vectors were chosen to steer retrieval is still a valid signature.

\subsection{Floating-Point and Quantised Numerics}\label{sec:bg-numerics}
Recomputing a vector does not, in general, reproduce it bit for bit. Floating-point addition is not associative~\cite{goldberg1991}, so the result of a reduction depends on its order, and the order depends on kernels, batch sizes, padding, thread counts and hardware. Large-language-model serving has recently been shown to be nondeterministic for exactly this reason, because kernels are not invariant to batch composition~\cite{thinkingmachines}. Quantisation adds a second, larger source of deviation~\cite{gholami2021}. Static quantisation fixes scales offline from calibration data; dynamic quantisation computes them at run time from the inputs, so that, with per-tensor scales, one input can change the computation applied to another in the same batch~\cite{quantamination}. Integer-only inference~\cite{jacob2018,ibert} avoids floating-point reductions altogether and can be made bit-identical across devices~\cite{detllm,intllm}.

Accelerators such as neural processing units execute quantised graphs produced by a vendor compiler; how far their outputs deviate from floating point, and whether they are bit-reproducible, can depend on the compiled graph and the toolchain. We measure these properties for floating-point and INT8 pipelines on general-purpose hardware (Sect.~\ref{sec:res-exact}) and leave dedicated accelerators to future work.

\subsection{Threat Model}\label{sec:threat}
\emph{System.} A publisher distributes a pack $P=\{(t_i,v_i)\}$ of text chunks $t_i$ and vectors $v_i$, signed under the pack's key hierarchy. A verifier, either the edge device itself or an auditor of a vector store, can compute $f(t_i)$, the embedding of $t_i$ under the agreed model and configuration. For a vector store, $v_i$ is replaced by stored codes $c_i$ under a codebook.

\emph{Adversary.} The adversary can make the verifier receive a pack or a store whose signatures are valid: a compromised publisher or build pipeline, a malicious insider who holds a content key, or a compromised writer of a vector store. Every text chunk the adversary ships passes content review, and the adversary chooses the vectors $v_i$ or the codes $c_i$. Besides tampering with the vectors of existing chunks, the adversary may add chunks whose text passes review, as a holder of a content key can; in the end-to-end experiment it adds one such chunk, an editorial note whose honest vector would not bring it into the context. The adversary knows the embedding model, which is public, and in the worst case the target queries; we also evaluate an adversary who knows only other questions about the target passage.

\emph{Goals.} \emph{Hijacking} places a chosen chunk at rank~1 for a target query instead of the gold chunk. \emph{Payload delivery} places a chosen chunk, for example one containing an instruction to the language model, in the top-$k$ context.

\emph{Verifiers.} We consider five verifier designs. (V0)~No verification. (V1)~Signature verification only. (V2)~Recomputation with tolerance: accept $v_i$ if $\cos(v_i,f(t_i))\ge\tau$. (V3)~Exact equality with a canonical computation: accept $v_i$ only if $v_i=f_c(t_i)$ bit for bit, where $f_c$ is a computation demonstrated to be bit-reproducible at publisher and verifier. (V4)~Code-level verification for compressed stores: accept the stored codes only if each entry equals the verifier's own code of $f(t_i)$, or equals its neighbouring code at an entry the verifier itself marks as ambiguous, and then replace the stored codes with the verifier's codes (Sect.~\ref{sec:res-exact}).

\emph{Assumptions and scope.} The verifier's embedding model, configuration, code and trust anchors are not under the adversary's control, and the codebook is obtained from a trusted source. We do not consider attacks that modify the model weights, the text of the chunks~\cite{r78,zhong2023}, the index structure beyond the stored vectors or codes, or the query channel~\cite{greshake2023}; nor side channels or denial of service.

\subsection{Related Work}\label{sec:related}

\emph{Poisoning and manipulation of retrieval.} Adversarial passages inserted into a corpus can be retrieved for many queries~\cite{zhong2023}, and crafted texts can make a RAG system produce attacker-chosen answers~\cite{r78}. Indirect prompt injection places instructions in content that an LLM-integrated application retrieves~\cite{greshake2023}. Certifiably robust aggregation bounds the influence of a small number of corrupted passages~\cite{r80}. These attacks change the \emph{text} that enters the corpus, and defences inspect or aggregate text. Our attack manipulates the vectors associated with fixed, review-admitted text, rather than optimising the text to change its embedding; content review and text-level defences do not see this manipulation. OWASP lists vector and embedding weaknesses among the top risks for LLM applications, including poisoning by insiders or unverified data providers, and recommends regularly auditing and validating the integrity of the knowledge base~\cite{owasp25}; it does not prescribe how stored vectors are to be bound to their text.

\emph{Attacks on and through embeddings.} Embeddings leak information about their inputs~\cite{song2020}, and text can be largely reconstructed from them~\cite{morris2023}. Closest to our setting, VectorSmuggle shows that an attacker with write access to an ingestion pipeline can hide data in embeddings with small perturbations that preserve retrieval, and proposes VectorPin, which signs each embedding together with its source content and producing model so that any later modification is detected~\cite{vectorsmuggle}. Provenance of this kind detects tampering \emph{after} signing. Our adversary is the signer, or a writer whose output is signed, and the question is whether the signed vectors are consistent with the text; we quantify what tolerance-based consistency checks concede and give checks that concede nothing under stated assumptions (a bit-reproducible computation, or a trusted codebook with canonicalisation).

\emph{Integrity and provenance in software and model supply chains.} Update frameworks, transparency logs and signing infrastructures protect the distribution of artefacts against key compromise, rollback and tampering~\cite{r85,ct,sigstore,intoto2019}, and model-signing specifications extend these guarantees to machine-learning models~\cite{r83}. They authenticate bytes and their history, not the semantic consistency between two parts of an artefact. Knowledge packs~\cite{prior} inherit these protections and, as we show, need an additional binding between text and vectors.

\emph{Determinism, reproducibility and verifiable inference.} Floating-point non-associativity~\cite{goldberg1991} and the lack of batch invariance~\cite{thinkingmachines} make inference results depend on how a request is executed; dynamic quantisation can even couple co-batched requests~\cite{quantamination}. Verifiable-inference systems handle this in two ways. Tolerance-based schemes accept results that are close to a recomputation, using locality-sensitive commitments to intermediate activations~\cite{toploc}. Exact schemes remove nondeterminism, either with pinned kernels and drivers on a fixed GPU architecture~\cite{eigenai} or with integer-only arithmetic that gives identical bits across different devices~\cite{detllm,intllm}, building on integer-only quantisation~\cite{jacob2018,ibert}. Quantisation can also be abused to hide malicious model behaviour that appears only after deployment~\cite{egashira2024}. Our work transfers the tolerance-versus-exactness question from generated text to retrieval: the margin law quantifies exactly what a tolerance concedes in vector search, and code-level verification shows that coarse quantisation cells make exact checks portable across floating-point platforms. We also measure the reproducibility of common floating-point and INT8 pipelines.

\emph{Vector search and quantisation.} Product quantisation~\cite{pq}, anisotropic quantisation~\cite{guo2020}, proximity graphs~\cite{r43} and GPU-scale libraries~\cite{faiss,johnson2021} are optimised for recall, memory and speed. VectorSmuggle also examines writer-side manipulation of quantised embeddings, with a different goal~\cite{vectorsmuggle}; to our knowledge, the retrieval-hijacking exposure of compressed vector stores under recomputation-based integrity audits, and its removal by verifier-controlled canonicalisation, have not been quantified.

\emph{Edge RAG, abstention and update screening.} Edge RAG systems reduce memory and latency on constrained devices~\cite{r16}. Selective answering under domain shift~\cite{r25} and unanswerability in RAG~\cite{r29} have been studied on server hardware, building on datasets with unanswerable questions~\cite{r105}. Natural language inference has been proposed to verify answers~\cite{r109}, but NLI models are known to rely on lexical overlap~\cite{hans}, which matters for screening edits that change a single fact. Our earlier study measured trust mechanisms on a CPU-only Raspberry Pi~5~\cite{prior}; this paper adds the text--vector binding and vector databases.

Table~\ref{tab:positioning} summarises how this work relates to the closest lines of research.

\begin{table}[!t]
\tablebodyfont\centering
\caption{Positioning of this work relative to the closest lines of research}\label{tab:positioning}
\begin{tabularx}{\textwidth}{p{0.22\textwidth}p{0.2\textwidth}XX}
\toprule
\TCH{Line of work} & \TCH{Representative} & \TCH{What it protects or shows} & \TCH{What it leaves open for our setting} \\
\midrule
Retrieval poisoning & \cite{zhong2023,r78,r80} & Attacks and defences for malicious \emph{text} in the corpus & Benign text with malicious vectors \\
Embedding provenance & \cite{vectorsmuggle} & Detects modification of a vector after it was signed & A compromised signer or writer \\
Supply-chain signing & \cite{r85,ct,sigstore,intoto2019,r83} & Authenticity and history of artefacts & Consistency between text and vectors inside an artefact \\
Tolerance-based verifiable inference & \cite{toploc} & Detects deviating LLM computations under float noise & What a tolerance concedes when outputs are retrieval vectors \\
Exact verifiable inference & \cite{eigenai,detllm,intllm} & Bit-identical results by pinning or integer arithmetic & Compressed stores and retrieval vectors \\
Vector quantisation & \cite{pq,guo2020,faiss,johnson2021} & Memory- and speed-efficient search & Integrity audits under quantisation error \\
\textbf{This work} & -- & Margin law for tolerance checks; exposure of compressed stores; code-level verification; on-device cost and impact & Edge accelerators; IVF and graph indexes; text-and-vector attackers \\
\botrule
\end{tabularx}
\end{table}

\section{The Margin Law}\label{sec:law}
Let $q$ be a normalised query vector, $g$ the gold unit (passage or window) and $C$ the set of other units, all normalised; when several units are relevant (positive BEIR qrels, or the windows of the gold SQuAD passage), $g$ is the closest of them and $C$ contains only non-relevant units. Let $\theta_g=\arccos(q\cdot g)$ and $\theta_c=\min_{c\in C}\arccos(q\cdot c)$, and consider queries where the gold unit ranks first, i.e.\ $\theta_g\le\theta_c$. In the cross-corpus analysis (Fig.~\ref{fig:law}) a gold unit tied with a competitor counts as first; in Table~\ref{tab:law}, ties are broken by index order. Ties give zero margins, which are hijackable at any non-zero tolerance. Define the \emph{angular margin} $m(q)=\theta_c-\theta_g$. A verifier of type V2 accepts a shipped vector $\tilde v$ for a chunk with honest vector $v$ whenever $\angle(\tilde v,v)\le\alpha$ with $\alpha=\arccos\tau$ (recomputation noise is discussed below). The optimal attacker rotates the closest non-gold vector towards $q$ by $\alpha$, which reduces its angle to $q$ to $\max(\theta_c-\alpha,0)$. For $\theta_g>0$, rank~1 is hijacked if and only if $\max(\theta_c-\alpha,0)<\theta_g$, that is $m(q)<\alpha$; if $\theta_g=0$ (a query identical to its gold vector), no competitor can rank strictly above the gold unit, and the law is an upper bound. Over a query set $Q$,
\begin{equation}
\mathrm{HR}(\tau)=\frac{1}{|Q|}\sum_{q\in Q}\mathbf{1}\left[m(q)<\arccos\tau\right],
\end{equation}
the fraction of margins strictly below $\arccos\tau$. An attacker who may also rotate the gold vector away by $\alpha$ obtains $m(q)<2\alpha$; we report the conservative single-rotation case. Five remarks delimit what the law measures. First, $\mathrm{HR}$ is \emph{per-query exposure}: the share of queries for which some admissible tampering of one vector hijacks rank~1. We do not claim that one tampered index realises all per-query hijacks at once; the end-to-end experiment of Sect.~\ref{sec:res-e2e} builds one tampered context per question. Second, it is conditional on queries whose gold unit ranks first under honest vectors, which is where hijacking is defined; on SQuAD this is 78\% (BGE-M3) and 70\% (MiniLM) of the answerable questions, and exposure over all questions scales by these factors. Third, at very small tolerances the exposed queries are those whose closest competitor is a near-duplicate of the gold unit; displacing the gold by such a competitor is not necessarily harmful, so small-tolerance exposure is an upper bound on harm, and harm is measured separately in Sect.~\ref{sec:res-e2e}. Fourth, we ignore recomputation noise. It is orders of magnitude smaller than all but the smallest tolerances we consider and shifts the accepted region by at most its own magnitude in either direction; exposure at a tolerance equal to the noise level (the FP32 rows of Table~\ref{tab:law}) is therefore an idealised figure. Fifth, the law asks whether some non-gold chunk can be made to outrank the gold unit, using the closest one; delivering a specific, fixed chunk is harder and is measured separately (Sect.~\ref{sec:res-e2e}). The law has two practical consequences. An operator can compute $\mathrm{HR}(\tau)$ from honest vectors and a query log, without attacking. And the safety of a tolerance depends on the corpus and the embedding model, not only on the tolerance. Numerically, recomputing all margins in double precision from normalised vectors with the stable formula $2\arcsin(\|q-d\|/2)$ changes no value by more than $4\cdot10^{-5}$~degrees and leaves every reported rate unchanged, for both the passage-level and the window-level units; margins that are exactly zero correspond to documents with identical text.

We also evaluate a \emph{transfer} attacker that does not know the user's wording: it rotates the carrier towards the centroid of other questions about the same passage, and success is measured on the real question.

\section{Experimental Setup}\label{sec:method}

\emph{Hardware.} Raspberry Pi~5 (16~GB, Arm Cortex-A76 CPU), on which the language model, the device-side embedding in ONNX Runtime and the pack verifier run on the CPU. A desktop (RTX~4090, 16 cores, WSL2) builds artefacts and runs scale experiments. An Apple MacBook Air (M2, 16~GB) is a third platform for the cross-device code-verification measurements.

\emph{Models.} Qwen2.5-3B-Instruct (Q4\_K\_M) under llama.cpp~\cite{r9,r104}, with the commit, prompts and settings of our earlier study~\cite{prior}, whose scoring we also reuse (an answer is correct if it is not a refusal and contains the normalised gold answer; refusals are detected with the earlier study's abstention rule); BGE-M3 dense retrieval~\cite{r42}; all-MiniLM-L6-v2~\cite{sbert,minilm} on 128-token windows, exported to ONNX and run in ONNX Runtime in FP32 and in dynamic INT8 (weights QInt8); ms-marco-MiniLM-L-6-v2 (cross-encoder), nli-MiniLM2-L6-H768 (natural language inference, NLI) and minilm-uncased-squad2 (extractive QA)~\cite{hfmodels}. Model revisions are pinned for all desktop and Raspberry Pi runs. The Apple M2 replication (Sect.~\ref{sec:res-exact}) loaded the current revisions and recorded their hashes (all-MiniLM-L6-v2 \texttt{1110a243}, BGE-M3 \texttt{5617a9f6}); both equal the pinned revisions.

\emph{Data.} The 950-passage SQuAD~2.0 corpus and 300 questions of~\cite{prior} (100 answerable, 100 in-topic unanswerable, 100 out of corpus)~\cite{r105}; for scale, all 4,690 answerable SQuAD~2.0 development questions on the corpus; four BEIR corpora (SciFact, NFCorpus, ArguAna, FiQA)~\cite{beir}; and 1,000 MS MARCO v2.1 queries~\cite{msmarco}.

\emph{Protocol.} Hypotheses, metrics and decision gates were written and dated before measurement and are released with the code; we call them pre-registered in this sense, as the protocol was not deposited in a public registry. Every deviation and exploratory analysis is logged with date and rationale (34 entries; the log released with the code omits the entries about experiments that are not reported in this paper). Reproducibility checks: the corpus is byte-identical to the earlier release, retrieval returns identical top-5 lists, and the earlier main configuration reproduces 60 of 60 answers exactly. Latencies are reported without llama-server prompt-cache hits, which we detect from the server's prompt timings.

\section{Results}\label{sec:results}

\subsection{RQ1: Tolerance Is an Attack Budget}\label{sec:res-law}
\begin{figure}[!t]
\centering
\includegraphics[width=0.85\textwidth]{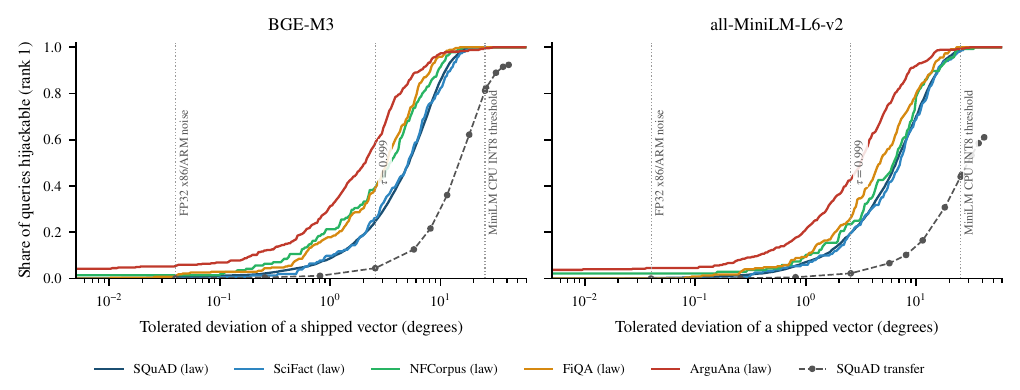}
\caption{Hijack rate as a function of the tolerated deviation, for SQuAD and four BEIR corpora (margin law, passage-level retrieval for both models) and the transfer attacker on SQuAD. Vertical lines mark FP32 cross-architecture noise, $\tau=0.999$ and the MiniLM CPU-INT8 reference threshold}\label{fig:law}
\end{figure}

On the 300-question set, the margin-law prediction equals the simulated attack at every tolerance (maximum absolute difference 0.0). Table~\ref{tab:law} reports the law on all answerable SQuAD questions whose gold unit ranks first: 3,648 with BGE-M3 on passages and 3,301 with MiniLM on 128-token windows, the unit the device embeds. BGE-M3 uses the passage index of the earlier study. Fig.~\ref{fig:law} and the comparison with the BEIR corpora instead encode every corpus, SQuAD included, at passage level with inputs truncated at 512 tokens for BGE-M3 and 256 for MiniLM, giving 3,646 and 3,148 queries. The two BGE-M3 encodings agree to within a cosine of $1-10^{-8}$ for the 940 passages of at most 512 tokens and differ only for the 10 longer ones; the two-query difference comes from questions about one such passage. For MiniLM the unit differs (passages instead of windows), and the exposure at $\tau=0.999$ is 19.7\% instead of 16.8\%. Median margins are 5.2\degr\ and 7.4\degr; 5\% of queries have margins below 0.6\degr\ and 0.8\degr. A tolerance of 0.999 (2.6\degr) leaves 25\% and 17\% of queries hijackable. As an INT8 reference we use the 0.1\% quantile (linear interpolation) of the cosines between dynamic-INT8 and FP32 outputs of all-MiniLM-L6-v2 in ONNX Runtime on the Raspberry Pi CPU over 1,745 windows, encoded with four threads in batches of 32 padded to the longest text in each batch (0.9040, 25.3\degr). Two of the 1,745 honest windows fall below it, the lowest at 0.898. Because dynamic quantisation depends on the batch, the threshold belongs to this configuration; re-encoding with the same settings reproduces the calibration vectors bit for bit, whereas batches of 64 change them by up to 0.03. It leaves 100\% and 98.4\% of queries hijackable, and 81.2\% and 44.0\% for an attacker who does not know the exact wording of the query. The threshold was calibrated for MiniLM; applying the same angular budget to BGE-M3 is a sensitivity analysis, not a measured BGE-M3 tolerance.

\begin{table}[!t]
\tablebodyfont\centering
\caption{Margin law on SQuAD~2.0: share of queries an attacker can hijack (rank~1) under a verifier with cosine tolerance $\tau$; BGE-M3 on passages (3,648 queries), MiniLM on 128-token windows (3,301 queries); 95\% Wilson intervals are within $\pm$1.8 points. ``Transfer'': attacker who knows the target passage and other questions about it, but not the exact wording of the query}\label{tab:law}
\begin{tabular}{lrcccc}
\toprule
\TCH{Tolerance $\tau$} & \TCH{Angle} & \TCH{BGE-M3 law} & \TCH{BGE-M3 transfer} & \TCH{MiniLM law} & \TCH{MiniLM transfer} \\
\midrule
1 (exact) & 0\degr & 0.00\% & 0.00\% & 0.00\% & 0.00\% \\
0.99999976 (FP32 x86 vs ARM) & 0.04\degr & 0.49\% & 0.05\% & 0.24\% & 0.03\% \\
0.99999 & 0.26\degr & 2.2\% & 0.5\% & 1.3\% & 0.15\% \\
0.999 & 2.56\degr & 24.6\% & 4.5\% & 16.8\% & 2.3\% \\
0.99 & 8.11\degr & 74.9\% & 21.7\% & 54.7\% & 10.2\% \\
0.95 & 18.19\degr & 99.6\% & 62.2\% & 91.9\% & 30.8\% \\
0.904 (MiniLM CPU INT8 reference) & 25.31\degr & 100\% & 81.2\% & 98.4\% & 44.0\% \\
0.8 & 36.87\degr & 100\% & 91.6\% & 100\% & 58.5\% \\
\botrule
\end{tabular}
\end{table}

Margins are small in every domain (Fig.~\ref{fig:law}). Median margins are 5.1--7.2\degr\ on SciFact, 3.5--6.3\degr\ on NFCorpus, 3.3--4.7\degr\ on FiQA and 2.1--3.2\degr\ on ArguAna, where 5\% of queries are decided by less than 0.02\degr\ (BGE-M3). With FP32 cross-architecture noise as the tolerance, 0--5.5\% of queries are exposed; with the MiniLM CPU-INT8 reference threshold, 98.7--100\% in all five corpora. Most of the small-tolerance exposure in ArguAna and NFCorpus comes from exact duplicates. Of the 23 (BGE-M3) and 17 (MiniLM) ArguAna queries exposed at the FP32-noise tolerance, 18 and 14 have a margin of exactly zero, and in every one of these cases the competing document's text is identical to the gold document's after collapsing whitespace and lower-casing, while only one of the two documents is marked relevant in the qrels; the same holds for the 2 exposed NFCorpus queries with BGE-M3 and 2 of 3 with MiniLM. Displacing a document by its own copy is harmless. Removing these cases from the exposed count (the denominator is unchanged), exposure at the FP32-noise tolerance is 5/420 (1.2\%) and 3/373 (0.8\%) on ArguAna, 0/141 and 1/136 on NFCorpus, 18/3,646 (0.49\%) and 7/3,148 (0.22\%) on SQuAD, 2/269 (0.74\%) and 0/225 on FiQA, and 0 on SciFact; these remaining cases are genuinely close, non-identical competitors. They are the non-zero intercepts of Fig.~\ref{fig:law}.

\subsection{RQ2: Compressed Vector Stores}\label{sec:res-vdb}
\begin{figure}[!t]
\centering
\includegraphics[width=0.95\textwidth]{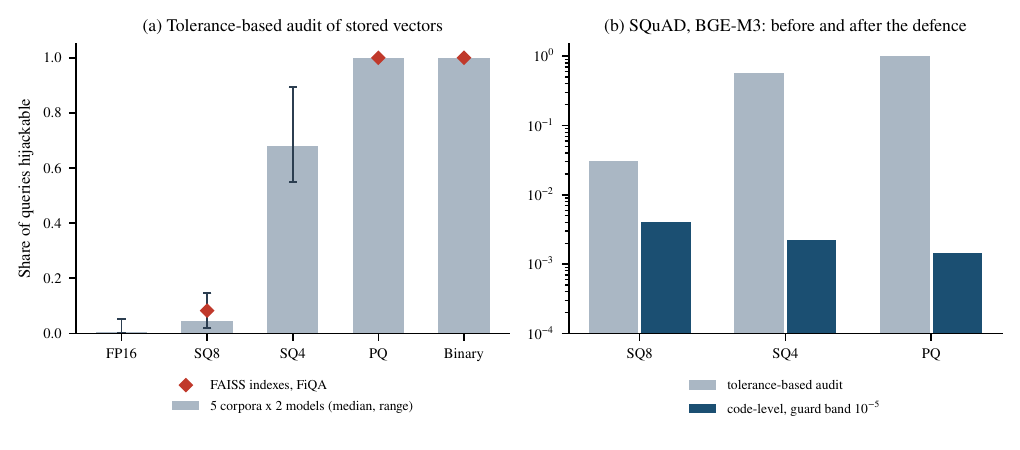}
\caption{(a) Share of queries hijackable under a tolerance-based audit of stored vectors, per storage format (bars: median and range over five corpora and two models; diamonds: FAISS indexes over FiQA). (b) The same exposure on SQuAD with BGE-M3 before (tolerance-based audit) and after the defence (code-level verification with a deliberately wide $10^{-5}$ guard band and without canonicalisation; at the calibrated widths the residual is at most 0.16\%, and canonicalisation removes it). Panel (b) shows the three formats covered by the residual analysis; FP16 stores no quantisation codes, and binary codes were not part of that analysis. Logarithmic scale}\label{fig:vdb}
\end{figure}

An audit that compares a recomputed embedding with a \emph{stored} vector must tolerate the storage format's own reconstruction error. Setting the tolerance at the largest honest reconstruction angle (no false alarms on our data) and applying the margin law gives Table~\ref{tab:vdb} and Fig.~\ref{fig:vdb}a. Half-precision storage has the lowest measured exposure; per-dimension 8-bit scalar quantisation leaves 2--15\% of queries exposed; product quantisation and binary codes leave essentially all queries exposed.

\begin{table}[!t]
\tablebodyfont\centering
\caption{Exposure of tolerance-based audits against compressed vector stores (5 corpora, 2 models; ranges over corpus--model pairs)}\label{tab:vdb}
\begin{tabular}{lccc}
\toprule
\TCH{Storage} & \TCH{Bits/dim} & \TCH{Max.\ honest reconstruction angle} & \TCH{Hijackable queries} \\
\midrule
FP16 & 16 & 0.03\degr & 0--5.2\% \\
SQ8 & 8 & 0.35--0.56\degr & 2.0--14.5\% \\
SQ4 & 4 & 5.8--9.6\degr & 55--89\% \\
PQ ($d/8$ subvectors, 256 centroids) & 1 & 23--40\degr & 99.7--100\% \\
Binary (sign) & 1 & 41--47\degr & 100\% \\
\botrule
\end{tabular}
\end{table}

\emph{Executed attack on a real vector database.} We repeated the analysis with FAISS~\cite{faiss} indexes over all 57,638 BEIR FiQA documents (all-MiniLM-L6-v2 with inputs truncated at 128 tokens, so the number of queries whose gold document ranks first, 230, differs from the 225 of the cross-corpus analysis at 256 tokens; publisher vectors on GPU, verifier recomputation on CPU), with the tolerance set at each index's largest honest reconstruction angle (Table~\ref{tab:faiss}). We then executed the attack against the product-quantisation index: for 60 target queries drawn at random (fixed seed) from those whose gold document ranks first under the exact cosine ranking of the uncompressed vectors, the attacker replaces the codes of a carrier document (the highest-ranked non-relevant document at ranks 11--60 of that same exact ranking) with the codebook entries that most raise its score, reverting the lowest-gain changes until the decoded vector passes the verifier's global tolerance check. FAISS itself is not at fault: it provides storage formats and search, not integrity audits, and serves here as a realistic implementation of the formats that such audits must tolerate. All 60 tampered vectors pass the check, and under the index's own scoring, the inner product of the query with the decoded PQ vectors of all 57,638 documents (the ranking \texttt{IndexPQ} computes with the inner-product metric), the carrier ranks first for 25\% of the targets (95\% CI 16--37\%). The margin-law columns of Table~\ref{tab:faiss} measure something different: the exposure of an angular tolerance audit, computed from the cosine margins of the uncompressed vectors with the tolerance set to the largest angle between each index's decoded vectors and the uncompressed ones; they do not depend on the index's search metric (Hamming distance for \texttt{IndexBinaryFlat}), and we do not claim that they bound the executed attack, which optimises a different objective. The executed attack is an existence result: a tampered store that passes the audit puts the attacker's document at rank~1 under the index's own scoring for at least a quarter of the targets. Code-level verification with canonicalisation (Sect.~\ref{sec:res-exact}) detects and removes all 60 tampered codes.

\begin{table}[!t]
\tablebodyfont\centering
\caption{Tolerance-based integrity checks on real FAISS indexes, BEIR FiQA (57,638 documents, 230 target queries with gold at rank~1). Tolerance: the largest angle between each index's decoded vectors and the uncompressed ones. Hijackable: margin law on the cosine margins of the uncompressed vectors, i.e.\ the exposure of the audit, independent of the index's search metric. The executed attack is scored with the index's own inner products of the query with the decoded PQ vectors}\label{tab:faiss}
\begin{tabular}{lcc}
\toprule
\TCH{FAISS index} & \TCH{Tolerance (largest honest angle)} & \TCH{Hijackable queries, margin law [95\% CI]} \\
\midrule
\texttt{IndexFlatIP} (float) & 0.00\degr & 0.0\% [0.0, 1.6] \\
\texttt{IndexScalarQuantizer} (SQ8) & 0.57\degr & 8.3\% [5.4, 12.5] \\
\texttt{IndexPQ} ($d/8\times$8 bits) & 37.97\degr & 100\% [98.4, 100] \\
\texttt{IndexBinaryFlat} & 44.64\degr & 100\% [98.4, 100] \\
\midrule
\multicolumn{3}{l}{Executed attack on \texttt{IndexPQ} (60 targets): 100\% of tampered vectors pass the tolerance check;} \\
\multicolumn{3}{l}{carrier ranks first 25\% [16, 37]; blocked by code-level verification with canonicalisation 100\% [94, 100]} \\
\botrule
\end{tabular}
\end{table}

\subsection{RQ3: Exact Verification}\label{sec:res-exact}

\emph{Determinism of inference pipelines.} Exact verification by recomputation requires bit-reproducible inference. PyTorch FP32 on GPU and CPU changes with batch size (maximum absolute coordinate difference about $10^{-7}$; FP16 on GPU $10^{-4}$), and so does PyTorch on an Apple M2 CPU and Metal GPU ($1.4\cdot10^{-7}$ and $4.5\cdot10^{-8}$), although repeated runs are identical; ONNX Runtime FP32 is batch and padding invariant on one x86 machine but differs between x86 and ARM ($1.9\cdot10^{-7}$); ONNX Runtime dynamic INT8 changes by up to 0.03 depending on the other inputs in the batch, which is the expected behaviour of per-tensor dynamic quantisation~\cite{quantamination}.

\emph{Code-level verification.} For compressed stores we propose to verify codes, not vectors. The verifier trusts only its own recomputation: it recomputes the embedding from the text, quantises it with the store's codebook and computes its own code $c$ for every entry. It marks an entry \emph{ambiguous} when, by its own recomputed value, honest float noise could have produced the neighbouring code $a$: for SQ when the value lies within $\varepsilon_{\mathrm{SQ}}$ of the rounding boundary (value units); for PQ when the squared distances to the nearest and second-nearest centroids differ by less than $\varepsilon_{\mathrm{PQ}}$ (squared-distance units, not comparable with $\varepsilon_{\mathrm{SQ}}$); for binary codes when $|v|<\varepsilon_{\mathrm{bin}}$. A stored vector is accepted only if every entry equals $c$, or is ambiguous and equals $a$; accepted vectors are then rewritten with the verifier's codes (canonicalisation). The publisher's float vectors are never used. We measured this verifier in both directions on three platforms (Table~\ref{tab:codes}), scanning guard widths. Exact equality fails rarely, and the smallest tested width that accepted every honest vector was the same on the desktop (NVIDIA GPU and x86 CPU) and on the Apple M2 (Metal GPU and ARM CPU): $3\cdot10^{-7}$ for MiniLM and $3\cdot10^{-6}$ for BGE-M3 with SQ8, and at most $3\cdot10^{-7}$, or none, for binary and PQ codes. For SQ and binary codes, whose guard widths are in value units, it was in every setting at or below the largest observed cross-device float deviation ($4.3\cdot10^{-7}$ and $1.0\cdot10^{-5}$ on the desktop, $3.6\cdot10^{-7}$ and $5.6\cdot10^{-6}$ on the M2). Guard exceptions are rare: on the Apple M2, at the smallest sufficient width, 0.9\% of MiniLM and 9\% of BGE-M3 SQ8 vectors relied on one to three exceptions. These widths are empirical calibrations on our samples and a grid of tested values, not guarantees for other data; a deployment should measure the deviation between its own publisher and verifier platforms.

\begin{table}[!t]
\tablebodyfont\centering
\caption{Code-level verification with the reference verifier (ambiguity decided by the verifier only; both directions). Each cell: share of honest vectors accepted with exact equality ($\varepsilon=0$, range over the two directions) / smallest tested guard width that accepted all vectors ($\varepsilon_{\mathrm{SQ}}$, $\varepsilon_{\mathrm{bin}}$ in value units; $\varepsilon_{\mathrm{PQ}}$ in squared-distance units). FiQA: MiniLM on 57,638 documents, BGE-M3 on a random sample (fixed seed) of 5,000; windows: 1,745 SQuAD windows, MiniLM in ONNX Runtime FP32 on x86 and on the Raspberry Pi's ARM CPU}\label{tab:codes}
\begin{tabularx}{\textwidth}{l>{\centering\arraybackslash}X>{\centering\arraybackslash}X>{\centering\arraybackslash}X>{\centering\arraybackslash}X>{\centering\arraybackslash}X}
\toprule
& \multicolumn{2}{c}{\TCH{Desktop: NVIDIA GPU $\leftrightarrow$ x86 CPU}} & \multicolumn{2}{c}{\TCH{Apple M2: Metal GPU $\leftrightarrow$ ARM CPU}} & \TCH{x86 $\leftrightarrow$ ARM (Pi)} \\
\TCH{Scheme} & \TCH{MiniLM} & \TCH{BGE-M3} & \TCH{MiniLM} & \TCH{BGE-M3} & \TCH{MiniLM windows} \\
\midrule
SQ8 & 98.99--99.03\% / $3\cdot10^{-7}$ & 91.3--91.5\% / $3\cdot10^{-6}$ & 99.12--99.13\% / $3\cdot10^{-7}$ & 91.0\% / $3\cdot10^{-6}$ & 99.26--99.43\% / $10^{-7}$ \\
SQ4 & 99.94\% / $10^{-7}$--$3\cdot10^{-7}$ & 99.5\% / $10^{-6}$--$3\cdot10^{-6}$ & 99.94--99.95\% / $10^{-7}$ & 99.3--99.4\% / $10^{-6}$--$3\cdot10^{-6}$ & 99.94--100\% / $\le10^{-7}$ \\
PQ & 99.995--99.997\% / $10^{-7}$ & 100\% / none & 99.995--99.997\% / $10^{-7}$ & 100\% / none & 100\% / none \\
Binary & 99.991\% / $10^{-7}$--$3\cdot10^{-7}$ & 99.94\% / $10^{-7}$ & 99.991\% / $10^{-7}$ & 99.94\% / $10^{-7}$ & 100\% / none \\
\midrule
Max.\ abs.\ coordinate difference & $4.3\cdot10^{-7}$ & $1.0\cdot10^{-5}$ & $3.6\cdot10^{-7}$ & $5.6\cdot10^{-6}$ & $1.9\cdot10^{-7}$ \\
\botrule
\end{tabularx}
\end{table}

The Apple M2 replication used the same corpus and the pinned model revisions; the x86--ARM comparison uses the windows embedded on the desktop and on the Pi (PyTorch on the GPU gives the same picture: largest deviation $2.6\cdot10^{-7}$, all vectors accepted at $10^{-7}$). The formats that make tolerance-based audits useless are thus the ones that make code-level audits most portable: coarse cells absorb float noise.

\emph{Residual budget and canonicalisation.} A compromised publisher could choose, at every ambiguous entry, the admissible code that raises a carrier and lowers the gold. We measured this on SQuAD with ambiguity decided by the verifier: at the calibrated widths, the publisher's choice hijacks no query for MiniLM (SQ8, SQ4 or PQ) and at most 0.16\% for BGE-M3 (SQ8 at $3\cdot10^{-6}$); with a needlessly wide band of $10^{-5}$ it reaches 0.41\%, against 2--100\% for tolerance-based audits. The residual disappears if the verifier \emph{canonicalises}: it accepts ambiguous entries but replaces them with its own codes, so the publisher has no choice left.

\subsection{RQ4a: Cost of Verification on the Device's CPU}\label{sec:res-cost}
Re-embedding the 1,745-window corpus on the Raspberry Pi takes 72.9~s in FP32 and 41.6~s in dynamic INT8 with four CPU cores (for reference, the earlier study's BGE-M3 indexing of the 950 passages on the CPU took 1,725~s). Running the embedding model on the CPU next to the language model is expensive (Table~\ref{tab:contention}): verifying 8 text--vector pairs per second costs the model 23--31\% of its decoding speed and raises the 99th-percentile inter-token latency by 108--167\%, and re-indexing at full speed costs 64--74\%. Verification on the device's CPU is therefore feasible but competes with generation; it is best scheduled when the model is idle or moved to a separate processor.

\begin{table}[!t]
\tablebodyfont\centering
\caption{Local LLM decoding on the Raspberry Pi~5 with an INT8 encoder (ONNX Runtime, 1--4 threads) running concurrently on the CPU: range over thread counts, three full runs (reference: 5.14 tokens/s, p99 inter-token latency 210~ms)}\label{tab:contention}
\begin{tabular}{lc}
\toprule
\TCH{Encoder load} & \TCH{Effect on the language model} \\
\midrule
Saturating (re-indexing, 15--25 items/s) & decode $-64.5\%$ to $-73.9\%$ \\
Periodic verification, 8 pairs/s & decode $-22.9\%$ to $-31.4\%$, p99 $+108\%$ to $+167\%$ \\
\botrule
\end{tabular}
\end{table}

\subsection{RQ4b: End to End on the Device}\label{sec:res-e2e}
\emph{Attack.} We built attacked contexts for the 100 answerable questions by adding one chunk with a fixed, instruction-bearing editorial note (stating that the requested information is not available in the collection), whose text is the same in every condition (the exact text and the construction are in the released code), and shipping, within each verifier's tolerance, a vector that places it in the top-3 context; we then ran the device's LLM. Without attack, the model answers 0.75 correctly and refuses 0.13. With an FP32 verifier at $\tau=0.999$, they are 0.74 and 0.13. Unlike the margin-law attacker, which rotates the closest competing passage, this attack must bring one fixed, off-topic instruction chunk into the context; a rotation of 2.6\degr\ is not enough for that, whereas the INT8 reference threshold (25.3\degr) is. With that threshold, correct answers fall to 0.39 and refusals rise to 0.50 (wrong answers 0.11). Delivery and effect separate cleanly: the payload entered the top-3 context for 100\% of questions without a vector check, 2\% at $\tau=0.999$ and 94\% at the INT8 reference threshold, and in every condition the shipped vector stayed within the angular budget and the same 100 questions were used.

\emph{Pack updates.} We built 19 update cases (12 structural attacks, three legitimate updates, three content-level attacks and one vector desynchronisation), verified on the device with FP32 vector checks; 17 of 19 decisions matched the expected outcome. Layered verification (signature, hashes, sequence and parent, declared change set, editor threshold for supersede and delete, revocation, transparency log) rejected all 12 structural attacks. The two mismatches concern content-level screening, which quarantines an added chunk when its maximum NLI contradiction probability against the nearest existing sentences reaches $\theta=0.5$: a benign addition was quarantined, and a poisoned addition that does not overlap existing content was not handled as expected (Sect.~\ref{sec:res-neg}). A naive vector desynchronisation (vectors far from their text) was quarantined; a tolerance-aware attacker, as shown above, passes a check at INT8-level tolerance.

\subsection{Negative Results}\label{sec:res-neg}
\emph{Abstention.} Relevance scores do not separate in-topic unanswerable from answerable questions on SQuAD (AUROC 0.69 bi-encoder, 0.63 cross-encoder), whereas an extractive-QA no-answer margin does (0.86). On MS MARCO the ranking reverses (bi-encoder 0.76, cross-encoder 0.83, QA margin 0.70): the QA margin, whose model was trained on SQuAD, is best on SQuAD, and the cross-encoder, trained on MS MARCO, is best on MS MARCO. The ranking of the scorers reversed between datasets, so no scorer can be assumed to transfer, and gates must be calibrated in domain.

\emph{Context pruning.} Keeping the three best sentences cut latency 2.9-fold (8.6~s versus 25.1~s) but correct answers fell from 0.75 to 0.44, although the answer string was present in 96\% of pruned contexts. Wrong answers on answerable questions barely changed (0.13 versus 0.12); instead the model abstained more on them (0.43 versus 0.13), and more often on in-topic unanswerable questions (0.80 versus 0.66), with out-of-corpus questions refused in both (0.99). Adding neighbouring sentences reduced abstention (0.30) but doubled wrong answers (0.25). The pre-registered non-inferiority criterion (a margin of 5 percentage points) is not met.

\emph{Semantic screening of updates.} An NLI screen for silent factual overrides failed (AUROC 0.18): the tested NLI model judges a near-copy with a changed fact as entailed, consistent with a known lexical-overlap bias~\cite{hans,r109}. An exploratory near-duplicate-edit signal reached 0.998 but is defeated by full paraphrases.

\section{Discussion}\label{sec:discussion}

\subsection{Why Tolerance Fails: the Geometry of Retrieval Margins}\label{sec:disc-geometry}
The central observation of this study is a mismatch of scales. Retrieval decisions in our corpora are taken by small angular differences: the median margin between the gold unit and its closest competitor is 2.1--7.4\degr\ across five corpora and two embedding models, and for 5\% of SQuAD queries it is below 0.6\degr\ (BGE-M3). The deviations that real pipelines introduce between a publisher's vector and a verifier's recomputation fall into three regimes that relate very differently to these margins.

\emph{Floating-point noise.} Recomputing the same model in FP32 on another device changes vector coordinates by $10^{-7}$--$10^{-5}$ (maximum absolute coordinate difference); the largest angular deviation we measured, between x86 and ARM, was 0.04\degr. This is far below typical margins, but not below all of them: queries whose closest competitor is a near-duplicate of the gold document have margins of the same order, which is why a tolerance at the level of cross-architecture noise still leaves up to 5.5\% of queries exposed (idealised margin-law exposure) (0\% in SciFact, most in ArguAna and NFCorpus), most of them because the corpus contains exact duplicates of the gold document, and up to 1.2\% otherwise. Exposure at this level is therefore a property of the corpus as much as of the tolerance, and its harm must be judged per query.

\emph{Quantised computation.} INT8 inference changes vectors by tens of degrees (median INT8--FP32 cosine 0.956 for all-MiniLM-L6-v2 in ONNX Runtime; the 0.1\% quantile, our reference threshold, corresponds to 25.3\degr). A threshold that accepts nearly all honest INT8 vectors is wider than almost every margin, and the margin law then predicts that 98--100\% of queries can be hijacked. An attacker who does not know the exact wording of the user's question, but knows the target passage and other questions about it, succeeds for 44--81\% of SQuAD queries.

\emph{Compressed storage.} Vector databases that store SQ, PQ or binary codes introduce reconstruction errors from a fraction of a degree (SQ8) to tens of degrees (PQ, binary). An audit that compares stored vectors with recomputed ones must tolerate these errors to avoid false alarms on honest data, and inherits the corresponding attack budget: 2--15\% for SQ8, 55--89\% for SQ4 and essentially all queries for PQ and binary codes. On real FAISS indexes the same pattern appears (8.3\% for SQ8, 100\% for PQ and binary), and an executed attack on a product-quantisation index placed the attacker's document first for a quarter of the targets with tampered vectors that all passed the check.

The practical consequence is that the safety of a tolerance cannot be judged in the abstract. It depends on the corpus, the embedding model and the storage format, and it can be measured. The margin law turns this measurement into a routine computation over honest vectors and a query log, without running any attack. The end-to-end experiment shows that the harm is concrete: on the device, an attack that stays within the INT8 reference threshold lowered correct answers from 0.75 to 0.39 and raised refusals from 0.13 to 0.50, by placing in the context an editorial note assumed to have passed content review.

\subsection{Choosing a Verification Strategy}\label{sec:disc-options}
Table~\ref{tab:options} compares the verification strategies we studied by what they guarantee, what they leave exposed in our measurements, what they require, and what they cost. Three observations follow.

\begin{table}[!t]
\tablebodyfont\centering
\caption{Verification strategies for shipped or stored vectors: guarantee, residual exposure measured in this study, requirements and cost}\label{tab:options}
\begin{tabularx}{\textwidth}{p{0.17\textwidth}XXX}
\toprule
\TCH{Strategy} & \TCH{Guarantee and residual exposure} & \TCH{Requirements} & \TCH{Cost} \\
\midrule
Signature only (V1) & Provenance of the bytes; no text--vector binding. A compromised signer controls retrieval. & Key management, transparency log & Negligible \\
Recompute and accept within $\tau$ (V2) & Exposure equals the margin law at $\arccos\tau$: 0--5.5\% at FP32 noise, 17--25\% at $\tau=0.999$ (SQuAD), 98--100\% at the MiniLM CPU-INT8 reference threshold & Same model and pre-processing as the publisher & One embedding per chunk \\
Recompute and discard & No exposure from shipped vectors: the device retrieves with its own vectors & A device-side embedder; the publisher's numerics become irrelevant & Full re-embedding (72.9~s in FP32, 41.6~s in INT8 for 1,745 windows on the Raspberry Pi CPU) \\
Exact equality with a bit-reproducible computation (V3) & No exposure & A computation shown to be bit-reproducible at publisher and device (for example identical hardware and software, or integer-only inference) & One embedding per chunk \\
Code-level verification without canonicalisation (V4 without its last step) & At calibrated guard widths, 0\% (MiniLM) to 0.16\% (BGE-M3, SQ8) & Trusted codebook and embedding configuration; calibrated guard width per platform pair & Re-embedding plus quantisation \\
Code-level verification with canonicalisation (V4) & No residual exposure by construction; all honest vectors accepted in our samples & As above & As above \\
\botrule
\end{tabularx}
\end{table}

First, the strongest guarantees come from removing the publisher's vectors from the trust base altogether, either by recomputing and discarding them on the device or by canonicalising stored codes. In both cases the verifier's own computation decides retrieval, and the publisher's choices of shipped vectors or codes have no effect on it. Second, exact equality is attractive because it needs no calibration, but it moves the trust problem to the reproducibility of the agreed computation: publisher and device must run a computation shown to be bit-reproducible, which none of the floating-point or dynamic-INT8 pipelines we measured is. Third, tolerance-based recomputation (V2) is the weakest option that still costs a full embedding pass; it only makes sense when its exposure, measured with the margin law on the deployment's own corpus, is acceptable.

A natural question is why a verifier that already recomputes embeddings should check the stored codes at all, rather than simply replace them. Replacement alone gives the same retrieval result, but it is silent: a store rewritten with tampered codes would be repaired without anyone learning that a writer had tampered with it. Verification before canonicalisation turns the inconsistency into an event: the pack or the writer can be rejected, the discrepancy logged, and update policy applied. The guarantee against steering comes from canonicalisation; the detection of out-of-policy inconsistencies, their logging and the policy come from the check, while attributing them to a party relies on the signatures. Both are bounded by the same assumptions: a trusted codebook and embedding configuration, a correct binding between document identifiers, text and codes, and protection of the store after the audit. Within those assumptions the publisher has no residual choice; outside them, for instance through the index structure, other controls are needed (Sect.~\ref{sec:disc-vdb}).

The cost of the stronger options is the cost of embedding. On the Raspberry Pi CPU, the full corpus is re-embedded in about a minute, and verification at 8 pairs per second alongside generation costs the language model 23--31\% of its decoding speed; a separate accelerator could reduce this cost, which we leave to future work. On larger hardware it is modest; for orientation, recomputing all-MiniLM-L6-v2 embeddings for the 57,638 FiQA documents took 277~s on the Apple M2 CPU and 106~s on its GPU, and BGE-M3 embeddings for 5,000 documents took 966~s and 462~s. Code-level verification can therefore run as a periodic or ingestion-time audit rather than on the query path.

\subsection{Design Guidelines}\label{sec:disc-guidelines}
We distil the results into recommendations for builders of offline RAG systems and vector stores.
\begin{enumerate}
\item \textbf{Do not treat a signature as a binding between text and vectors.} Signatures establish who published a pack; they do not show that its vectors encode its text. A compromised or malicious signer is exactly the case the binding must cover.
\item \textbf{Measure before choosing a tolerance.} Any verification tolerance is an attack budget. Compute the margin distribution of the deployment's corpus and model on representative queries and read the exposure off the margin law; pay attention to near-duplicates, which create exposure even at float-noise tolerances.
\item \textbf{Prefer recompute-and-discard on devices that can afford it.} If the device embeds the content itself and retrieves with its own vectors, the shipped vectors play no role in retrieval. On a small board the cost is about a minute per corpus of this size on the CPU.
\item \textbf{Audit compressed stores at the level of codes, and canonicalise.} Decide ambiguity from the verifier's own recomputation, accept a neighbouring code only where the verifier sees ambiguity, and replace accepted ambiguous entries with the verifier's codes. Calibrate the guard width from the measured float deviation between the actual publisher and verifier platforms, and recalibrate after any change of model, runtime or hardware.
\item \textbf{Bind the codebook and the configuration.} Code-level verification is only as trustworthy as the codebook and the embedding configuration it uses. Their hashes, together with the model revision, tokeniser, truncation length, pooling and normalisation, belong in the signed manifest; a verifier should never accept a codebook from the same party whose codes it verifies.
\item \textbf{Pin everything that affects numerics.} Model revisions, runtime and compiler versions and compiled accelerator graphs should be pinned and recorded, so that a verifier can reproduce the publisher's computation and an auditor can reproduce the verifier's.
\item \textbf{Require a demonstrably bit-reproducible computation for exact verification} at publisher and device.
\item \textbf{Schedule verification away from the language model's compute.} On the CPU of an edge device, verification at a modest rate slows decoding by roughly a quarter to a third; schedule it when the model is idle or move it to a separate processor.
\item \textbf{Calibrate abstention in domain and evaluate context pruning by its error decomposition}, not by accuracy alone.
\end{enumerate}

\subsection{Implications for Vector Databases and Integrity Guidance}\label{sec:disc-vdb}
Current guidance for LLM applications recommends regularly auditing and validating the integrity of the knowledge base and accepting data only from trusted sources~\cite{owasp25}, and per-embedding signatures have been proposed to make tampering after embedding detectable~\cite{vectorsmuggle}. Our results add two points. Provenance controls do not cover a compromised writer, who can sign or store whatever vectors it wants; and integrity audits that compare stored vectors with recomputed ones are ineffective for compressed formats that many production stores use. Code-level verification gives the recommended audit a concrete, measurable form that works for those formats, and it complements provenance: signatures tell the verifier who wrote a code, recomputation tells it whether the code is consistent with the text.

Code-level verification fits naturally into three points of a vector database's life cycle: at ingestion, where a writer's codes can be checked against an independent recomputation before they become searchable; at replication or synchronisation between sites, where a replica can verify the codes it receives; and in periodic background audits, which can be scheduled at low priority because the cost is that of re-embedding. Canonicalisation means that an audit does not merely detect inconsistency but repairs the ambiguous entries; in our samples and at the guard widths we tested, it rejected no honest vector.

Our FAISS experiments used flat indexes. Production systems often add a coarse quantiser (IVF) or a proximity graph (HNSW) on top of compressed codes. A coarse-quantiser assignment is itself a code and can be verified the same way; graph edges are not determined by the text alone and their integrity requires different mechanisms. We leave these structures to future work.

\subsection{Implications for Edge Devices}\label{sec:disc-hw}
Two properties of edge inference matter for trust mechanisms. First, determinism becomes a security property: exact verification is possible only if inference is bit-reproducible, and none of the pipelines we measured is (PyTorch on NVIDIA and Apple GPUs and on CPUs varied with batch size, ONNX Runtime FP32 varied across architectures, and dynamic INT8 varied with the other inputs in the batch). Stacks that are bit-reproducible by construction, such as integer-only inference~\cite{detllm,intllm} or pinned kernels on fixed hardware~\cite{eigenai}, are therefore the natural basis for exact verification at the edge. Second, cost: on the device's CPU, verification competes directly with the language model for compute. Dedicated accelerators may change both properties; measuring their determinism and their cost next to a running model is an important next step.

\subsection{Lessons from the Negative Results}\label{sec:disc-neg}
Three pre-registered expectations failed, and each carries a lesson for trust mechanisms in offline RAG. Relevance scores did not identify in-topic unanswerable questions, and the ranking of the abstention scorers reversed between SQuAD and MS MARCO, so none can be assumed to transfer. An offline device that must abstain safely therefore needs abstention thresholds calibrated on its own domain, which in a disconnected setting means shipping calibration data or thresholds with the knowledge pack. Context pruning reduced latency nearly threefold with almost no change in wrong answers, but it converted correct answers into abstentions; whether that trade is acceptable depends on the relative cost of a wrong answer and an abstention, and can only be judged from the full error decomposition. Finally, a small NLI model was not able to detect silent factual overrides, because it judges a near-copy with a changed fact as consistent. Structural governance, i.e.\ editor thresholds for changes to existing content, rejected all twelve structural attacks we tested and remains the reliable control for updates; semantic screening is at best a secondary signal.

\subsection{Threats to Validity}\label{sec:disc-threats}
\emph{Internal validity.} All code-level results were produced by a single reference verifier whose rules match the description in Sect.~\ref{sec:res-exact} and that is covered by unit tests. During the study, an earlier measurement that decided ambiguity from the publisher's vectors was superseded, and a first run of the reference verifier on one platform produced inconsistent counters that disappeared after a fix for constant dimensions and a change of numerical library version; only the corrected runs are reported, and both events are recorded in the protocol. Latencies of the language model are reported without prompt-cache hits, which we detected and removed.

\emph{Construct validity.} We measure hijacking as displacement of the gold unit from rank~1 (in the FAISS experiment, as the attacker's document ranking first under the index's scoring, the honest ranking being taken from the uncompressed vectors) and payload delivery as entry into the top-3 context; other definitions of harm are possible. The margin law describes an optimal attacker that rotates one competing vector; an attacker who can also move the gold vector has a larger budget, so our figures are conservative in that respect. Tolerances were set from honest data (the 0.1\% quantile of honest INT8 cosines, the largest honest reconstruction angle of a store); a deployment that accepts occasional false alarms could use tighter values, at the price of rejecting honest content. The executed FAISS attack is greedy and restricted to the codebook and is therefore a lower bound.

\emph{External validity.} We studied two embedding models, English corpora of up to 57,638 documents, flat FAISS indexes, one board with inference on its CPU, and one local language model. The CPU-INT8 reference threshold was measured for one model with one runtime and dynamic quantisation, whose outputs depend on the batch; other quantisation schemes and accelerators will have other tolerances, which the margin law turns into exposure. Guard widths are empirical calibrations from three platforms and a grid of tested values, not guarantees for other models or hardware. The update scenarios are synthetic.

\emph{Statistical conclusion validity.} Where confidence intervals are given for proportions they are Wilson intervals, the main systems result is the mean of three full runs with narrow ranges, and the end-to-end experiment uses 100 questions per condition. Exploratory analyses are labelled as such and logged with dates in the protocol.

\subsection{Ethics and Responsible Disclosure}\label{sec:disc-ethics}
The attacks studied here require write access to a knowledge pack or a vector store; they model compromised publishers, build pipelines and insiders rather than remote attackers. We present every attack together with a method to measure exposure and with defences, and we release the measurement tools so that operators can assess their own systems.

\subsection{Future Work}\label{sec:disc-future}
Several directions follow directly from the limitations above: measuring the determinism and cost of verification on edge accelerators; extending code-level verification to IVF and graph-based indexes and to multilingual corpora at larger scale; studying attackers who modify both text and vectors; deriving guard widths from a model of floating-point error rather than from measurement alone; and specifying a canonical, signed representation of embedding configurations and codebooks that verifiers in different systems could share.

\section{Conclusion}\label{sec:conclusion}
Offline retrieval-augmented generation depends on knowledge that arrives in packs, and packs carry vectors as well as text. A signature tells a device who published those vectors; it does not tell it whether they encode the text. We showed that any check which tries to establish that binding within a tolerance concedes a predictable share of queries to an attacker who controls the vectors, and that this share is given by the distribution of retrieval margins. Margins are small in every corpus we studied, so float-noise tolerances leave a measurable exposure, a threshold calibrated on INT8 inference leaves 98--100\% of queries exposed, and integrity audits of compressed vector stores leave up to all queries exposed. On a real FAISS index, an attack whose tampered vectors all passed such an audit ranked the attacker's document first under the index's own scoring for a quarter of the targets; on the device, a comparable attack nearly halved the language model's correct answers (from 0.75 to 0.39).

The gap can be closed. Recomputing vectors on the device and discarding the shipped ones removes the publisher's influence through the shipped vectors; exact verification removes it as well, provided that publisher and device share a bit-reproducible stack; and for compressed stores, verifying quantisation codes, with ambiguity decided by the verifier, a calibrated guard band and canonicalisation, accepted every honest vector in our samples on three platforms and four compute backends while leaving no choice to the publisher. On the CPU of a small board, the necessary recomputation is affordable but competes with the language model, which makes scheduling, or a separate processor, part of the design.

These results turn a recommendation that knowledge bases should be checked for integrity into a procedure that can be measured, calibrated and audited. We release the code, the data, the reference verifier and the protocol with its deviation log, so that operators can measure the exposure of their own systems and adopt the defences.

\section*{Statements and Declarations}

\bmhead{Use of AI tools}
The authors used standard writing and production tools (Grammarly, Overleaf) for language editing and \LaTeX\ preparation. Generative AI assistants were used to help identify candidate literature, to cross-check bibliographic metadata, and to help write the experiment and analysis scripts; every reference was verified by the authors against the publisher, DOI, or preprint-server record, and all code and results were reviewed by the authors. The AI tools did not generate the framework, the findings, or the conclusions. The authors take full responsibility for the manuscript.

\bmhead{Funding}
This work was funded by securagen.ai.

\bmhead{Competing interests}
The authors are affiliated with securagen.ai P.C., which develops commercial software in the areas discussed in this paper. The experiments used open-source software and publicly available model weights under their respective licences. The authors have no other relevant financial or non-financial interests to disclose.

\bmhead{Ethics approval and consent to participate}
Not applicable.

\bmhead{Consent for publication}
Not applicable.

\bmhead{Data availability}
Per-query results, vectors where licences permit, system information and analysis outputs are available at \repourl\ (\zenodourl). SQuAD~2.0, BEIR and MS MARCO are publicly available~\cite{r105,beir,msmarco}.

\bmhead{Code availability}
All scripts, the pre-registered protocol with its deviation log, and the pinned versions of models and tools are available at \repourl\ (\zenodourl). The implementation uses open-source software (llama.cpp, ONNX Runtime, sentence-transformers, FAISS). The generated answers released with the code were produced with Qwen2.5-3B-Instruct, used for research under the Qwen Research License; the embedding, reranking, NLI and QA models are used under their respective licences, and SQuAD 2.0, BEIR and MS MARCO under their terms of use. Raspberry Pi is a trademark of Raspberry Pi Ltd; other product names are trademarks of their respective owners.

\bmhead{Author contribution}
Georgios Feretzakis and Alexandros Papaspyridis had the idea for the article. Georgios Feretzakis designed the methodology, designed and conducted the experiments, analysed the results, wrote the manuscript, and supervised and administered the project. Alexandros Papaspyridis critically revised the work. Both authors read and approved the final manuscript.

{\raggedright\setlength{\parskip}{0pt}

}

\end{document}